\documentclass[]{emulateapj}
\usepackage{graphicx}
\usepackage{graphics}
\usepackage{amsmath}
\usepackage{color}
\usepackage{dashrule}
\usepackage{multirow,bigstrut,bigdelim}
\usepackage{epsfig}

\def\ep{\epsilon}

\def\G{\Gamma}
\def\g{\gamma}
\def\jcap{Jour. Cosmology and Astro-Particle Phys.} \def\mnras{M.N.R.A.S}
\def\apj{Astrophys.J.} 
 \def\nat{Nature}    \def\prl{Phys. Rev. Lett.}      \def\prd{Phys.Rev.D}

\begin{document}

\title{On the neutrino non-detection of GRB 130427A}

\author{Shan Gao} \email[Email(SG):]{ sxg324@psu.edu}
\author{Kazumi Kashiyama} \email[Email(KK):]{ kzk15@psu.edu} 
\author{Peter M\'esz\'aros~} \email[Email(PM):]{ pmeszaros@astro.psu.edu}
\affiliation{Department of Physics, Department of Astronomy and Astrophysics,\\
Center for Particle Astrophysics, The Pennsylvania State University, University
Park, 16802, USA} 

\date{\today} % delete this line to display the current date

\begin{abstract}

The recent gamma-ray burst GRB 130427A has an isotropic electromagnetic
energy $E^{iso}\sim10^{54}$ erg, suggesting an ample supply of target
photons for photo-hadronic interactions, which at its low redshift of $z\sim0.34$
would appear to make it a promising candidate for neutrino detection. However, the
IceCube collaboration has reported a null result based on a search during the
prompt emission phase. We show that this neutrino non-detection can provide valuable
information about this GRB's key physical parameters such as the emission radius 
$R_{d}$, the bulk Lorentz factor $\Gamma$ and the energy fraction converted into
cosmic rays $\epsilon_{p}$.  The results are discussed both in a model-independent 
way and in the specific scenarios of an internal shock model (IS), a baryonic 
photospheric model (BPH) and magnetic photospheric model (MPH).  We find that the
constraints are most stringent for the magnetic photospheric model considered, but 
the constraints on the internal shock and the baryonic photosphere models are
fairly modest.

\end{abstract}

\section{Introduction}
\label{sec:1}

Gamma-ray bursts (GRBs) have been proposed as a major source of high energy cosmic 
rays,  provided that a substantial fraction of protons are accelerated in the
inferred shocks or magnetic reconnection regions.  However, the underlying
mechanism of the prompt gamma-ray emission, the jet structure and the particle 
acceleration details remain uncertain. Very high energy neutrinos, however, 
would be a natural by-product from high energy protons interacting with other 
baryons or with photons, suffering little from absorption effect along the 
propagation path and providing valuable clues about the presence of cosmic rays.
It is expected that if a major fraction of the GRB energy is converted into 
ultra-high energy cosmic rays, a detectable neutrino fluence should appear in
IceCube \citep{Ahlers+11-grbprob}.  However, the two-year data gathered by the IceCube 
40 + 59 string configuration has challenged this scenario by a null
result in the search for correlation with hundreds of electromagnetically
detected GRBs \citep{Abbasi+12grbnu-nat}. 
Constraints on the conventional internal shock fireball 
models have been derived \citep{He+12grbnu} and several alternative models 
have been discussed \citep{Vurm+11phot,Zhang+11icmart,Gao+12photnu}. 

\begin{figure} % ============= Fig 1 ===================

\includegraphics[scale=0.8]{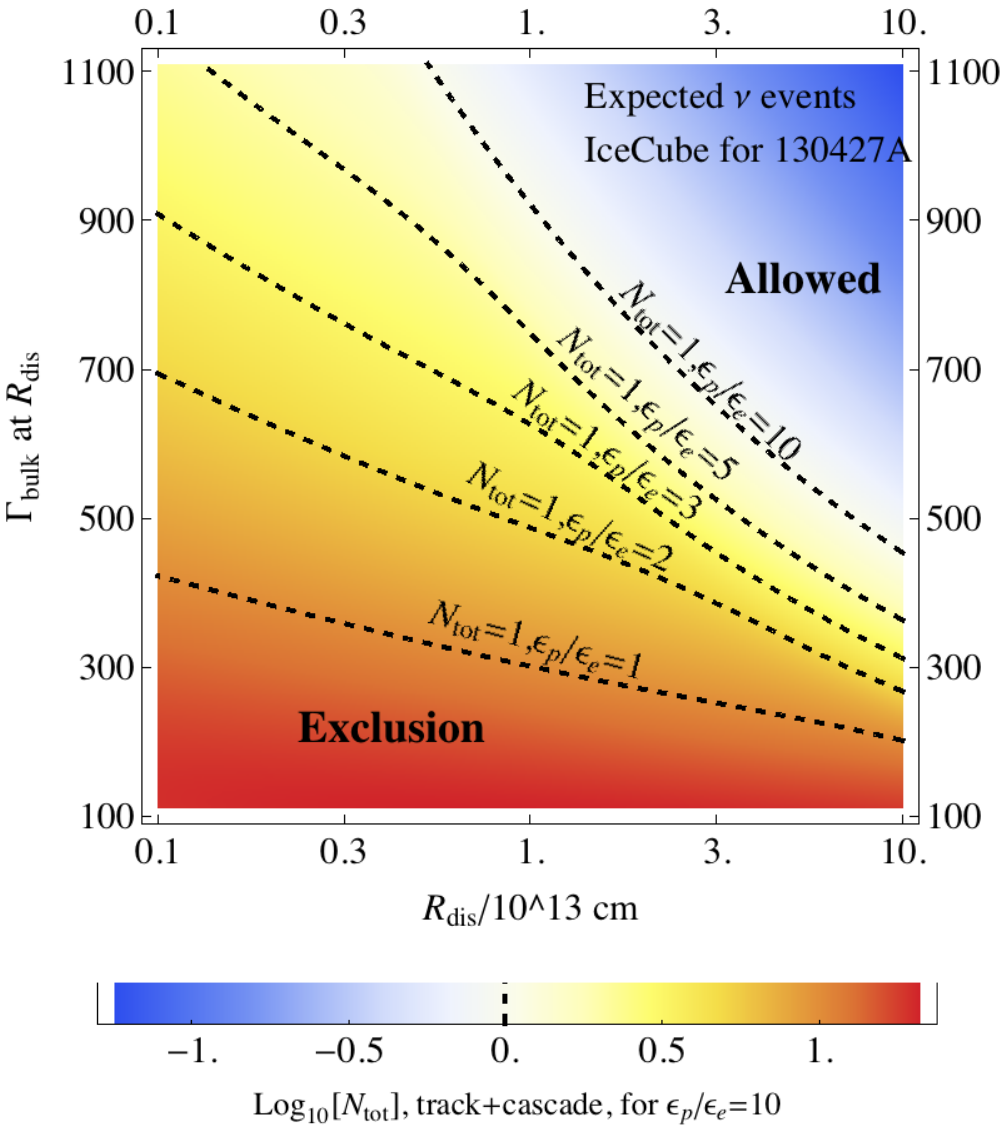}

\caption{\label{fig:F1} (See e-print for colored version) Density plot of the 
expected number of neutrino events (track+cascade) in IceCube for GRB 130427A 
on the 2D parameter space of the dissipation radius $R_{13}=R_{d}/10^{13}{\rm ~cm}$ 
and the bulk Lorentz factor $\G$ of the jet at this radius. This calculation uses
the semi-analytical method similar to \citep{Waxman+97grbnu,Zhang+12grbnu} but 
assuming no specific scenario (e.g. neither an internal shock, nor other model,
see \S II for details). The blue color (top-right region) denotes fewer events 
while the red (lower regions) denotes more events. The five dashed lines from 
top to bottom show contours where one event is expected, for different proton 
to electron energy ratios $\ep_{p}/\ep_{e} =10,5,3,2,1$. The other two energy 
partition parameters are taken to be constants, $\ep_{e}=0.1$ and $\ep_{B}=0.01$. 
Based on the null result in the IceCube neutrino search reported in 
\citep{Abbasi+12grbnu-nat}, 
the parameter space below each contours is more likely to be ruled out for the
corresponding $\ep_{p}/\ep_{e}$.}

\end{figure}
 
Recently a super-luminous
burst, GRB 130427A, was detected simultaneously by five different satellites, 
with an isotropic equivalent  energy of $E^{iso}\sim10^{54}$ ergs in gamma-rays at
a low redshift of $z\sim0.34$ \citep{Fan130427A}.  
Disappointingly, a  neutrino search for this GRB 
reported by the IceCube collaboration yielded a null result
\footnote{http://gcn.gsfc.nasa.gov/gcn3/14520.gcn3}. Here we show that this 
null detection is not surprising, and show that it provides interesting  information 
about the properties of this GRB, some of which are otherwise difficult to obtain 
through conventional electromagnetic channels.  We discuss the constraints on the
physical parameters of this GRB, both (a) using a model-independent procedure patterned
after that of \citep{Waxman+97grbnu} and \citep{Zhang+12grbnu}  
(section \ref{sec:2}), and  (b) for three specific GRB models (IS, BPH and MPH, 
section \ref{sec:3}), summarizing our results in section \ref{sec:4}). 

\section{Model-independent constraints on the dissipation radius, bulk Lorentz 
factor and total cosmic ray energy} \label{sec:2}

We assume a simple GRB jet model, whose Lorentz factor averaged over the jet cross 
has a value $\G$ at the dissipation radius $R_{d}$. Generally $R_{d}$ is model
dependent and is a function of $\G$. However, in the interest of generality,
in this section we do not specify the underlying models, leaving this for section \ref{sec:3}.  
Here, the parameters $\G$ and $R_{d}$ are treated as two independent variables.  
At $R_{d}$, a fraction of jet total energy $E_{tot}$ (in the form of kinetic energy 
and a possible toroidal magnetic field energy if the jet is highly magnetized) is 
dissipated and converted into energy carried by accelerated cosmic rays protons 
$\ep_{p}E_{tot}$, turbulent magnetic fields $\ep_{B}E_{tot}$ and high energy non-thermal 
electrons $\ep_{e}E_{tot}$ (the latter promptly converting into photons).

For GRB 130427A, the photon spectrum is well fitted by a Band-function spectrum
with $dN/dE\propto(E/E_{{\g}b})^{-s}$ in the observer frame, where $s=0.79$ for
$E<E_{{\g}b}$ and $s=3.06$ for $E_{\g{b}}<E<10$ MeV, with a spectral break
energy $E_{\g{b}}=1.25$ MeV and a total isotropic equivalent 
$E_{\g}\sim1.05\times10^{54}$ ergs \citep{Fan130427A}. For simplicity, in 
this paper we assume $\ep_{e}=0.1$, corresponding to a jet total energy, 
$E_{tot}=10E_{\g}$. The value of $\ep_{B}$ is uncertain 
see e.g. \citep{Lemoine+13}; here we use a value 
$\ep_{B}=0.01$. A high magnetic field would suppress the 
neutrino spectrum at very high energies, since the $\pi^{\pm}$ and $\mu^{\pm}$ would
have time to cool by synchrotron emission before they decay, see e.g. \citep{Rachen+98crnu}.
However for GRB 130427A, 
as we show below, the neutrino flux decreases rapidly as the energy increases above 
the peak, the final expected event rate in IceCube being insensitive to $\ep_{B}$. 
For the neutrino calculation, we follow the outlines in \citep{Waxman+97grbnu}, 
and \citep{Zhang+12grbnu}; see also 
\citep{Zhang+12grbnu,Li12ic3nu,He+12grbnu,Hummer+11nu-ic3} for detailed treatments.
For this specific GRB, the analytical approximations in this section 
lead to an error of $<30\%$ compared to the results obtained with the methods in section
\ref{sec:3}, for most of the realistic parameters.
The cosmic rays are accelerated 
to a $dN_{p}/dE \propto E^{-2}$ spectrum up to a maximum energy $E_{p,max}$ determined 
either by the Hillas condition or by $t_{dyn}=t_{acc}$ (in the jet frame and converted
to the observer frame), whichever is smaller. Here $t_{dyn}\sim{R_{d}}/\G{c}$ is the 
dynamical time scale and $t_{acc}\sim{o(1)}\times{E_{p}}/eBc$ is the acceleration 
time scale in the jet frame. High energy protons lose energy in the jet frame 
due to $p\g$ interaction, at a rate 

\begin{equation}
t_{\pi}^{-1}\equiv-\frac{d\g_{p}}{\g_{p}dt}=\frac{c}{2\g_{p}^{2}}
\int_{E_{TH}}^{\infty}dE\sigma_{p\g}\kappa E\int_{E/2\g_{p}}^{\infty}x^{-2}n(x)dx
\end{equation}
The second integral can be solved analytically for the
broken-power-law photon energy distribution $n(E)=dN_{ph}/dE$ , while for the
first integral is approximated using the $\Delta^{1232}$ resonance, in which,
$\sigma_{p\g}=\delta(E-E_{pk})5\times10^{-28}{\rm cm^{2}}$ where
$E_{pk}=0.3$ GeV is the peak of the $p\g$ cross section, $E_{TH}=0.2$ GeV is the
threshold and $\Delta{E}=0.3$ GeV is the width of the resonance measured in the 
proton rest frame. $\kappa=0.2$ is the averaged inelasticity of the $p\g$ interaction, or
$E_{p}=5E_{\pi}$. It is convenient to define the pionization efficiency 

\begin{equation}
f_{\pi}\approx min(1,t_{\pi}^{-1}/t_{dyn}^{-1})
\end{equation}
which describes the fraction of energy flowing from parent protons to pions within
the dynamical time scale. Of $f_{\pi}$, about $1/2$ goes to $\pi^{+}$ and
$1/2$ to $\pi^{0}$, and neutrinos are produced by the charged pion decay

\begin{equation}
\pi^{\pm}\rightarrow\mu^{\pm}+\nu_{\mu}(\bar{\nu}_{\mu})\rightarrow
e^{\pm}+\nu_{e}(\bar{\nu}_{e})+\nu_{\mu}+\bar{\nu}_{\mu}~.  
\end{equation}
The energy of the charged pion is approximately equally divided among the four
leptons, $E_{\pi}=4E_{\nu}$. Due to neutrino oscillations and the large
uncertainties in the exact energy and distance of the source, we assume that they 
arrive at the earth in equal numbers per flavor.  Thus, we can express the 
muon + anti-muon neutrino flux ($\phi\equiv{dN}/dE$) in the observer frame as 

\begin{eqnarray} 
\phi_{f_{\pi}<1}(E)=&
4.8\times10^{-12}L_{\g,53}^{2}(\ep_{p}/\ep_{e})(\ep_{\g{b}/{\rm MeV}})(1+z)^{2}
\nonumber \\ & \G_{300}^{-6}D_{27}^{-2}R_{13}^{-1}
(E/E_{\nu{b}})^{s-3}~~{\rm GeV^{-1}cm^{-2}}
\label{eq:phiA} \end{eqnarray}

\begin{eqnarray} 
\phi_{f_{\pi}\ge1}(E)=&
0.8\times10^{-12}L_{\g,53}(\ep_{p}/\ep_{e})(1+z)^{3}~~~~~~~~~~~~\nonumber \\ 
&\G_{300}^{-4}D_{27}^{-2}
 (E/E_{\nu{b}})^{-2}~~~~~~~~~{\rm GeV^{-1}cm^{-2}}	\label{eq:phiB}
\end{eqnarray}
where $s = 3.06$ (higher Band index) for $E<E_{\nu{b}}$ and
$s = 0.79$ (lower Band index) for $E>E_{\nu{b}}$. The pionization
efficiency in the above equation is

\begin{equation} 
f_{\pi} = 6.13L_{\g,53}R_{13}^{-1}(E_{\g{b}}/{\rm
MeV})\G_{300}^{-2}(1+z)^{-1}(E/E_{\nu{b}})^{\alpha-1} 
\end{equation}
with the first neutrino break energy (due to those protons interacting with the
$E=E_{\g{b}}$ photons)  at

\begin{equation} E_{\nu{b}} = 6.33\times10^{5}(E_{\g{b}}/{\rm
MeV})\G_{300}^{2}(1+z)^{-2}~{\rm GeV} \label{eq:nub} \end{equation}
and a second neutrino break energy (due to synchrotron and inverse Compton cooling by
charged pions, assuming Thomson regime for simplicity) \footnote{For energy
above this, it is good approximation to treat it as a cutoff for this GRB} at

\begin{eqnarray} E_{\nu{b},2} = &
2.12\times10^{7}L_{\g,53}^{-1/2}R_{13}\G_{300}^{2}(1+z)^{-1} \nonumber \\ &
(\ep_{B,-1}+{\ep_{e,-1}})^{1/2}~~{\rm GeV} \end{eqnarray}

% ========= Figure 2-4================== 

\begin{figure} \includegraphics[scale=0.8]{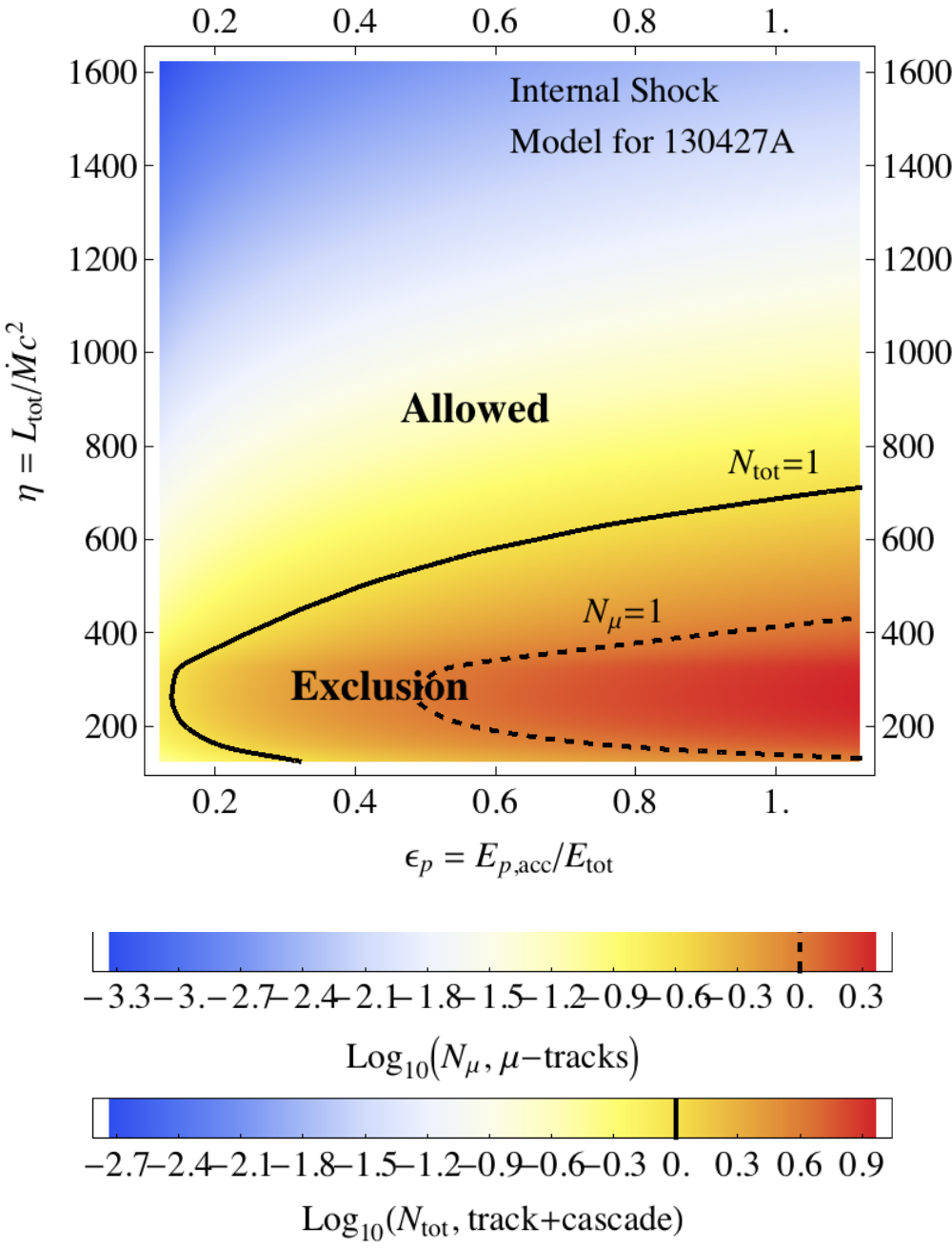}
\caption{\label{fig:F2} Density plot of expected neutrino event number in 
IceCube based on the internal shock model for GRB 130427A. The total number
$N_{tot}$ (muons + cascades) and $N_{\mu}$ (muons only) are calculated in the 
2D parameter space $\eta$ and $\ep_{p}$. 
Here $R_{d}$ is the internal shock radius for a 1 ms variability timescale 
(see \S III). The value of $N$ is represented 
the by color; however, for $N_{tot}$ and $N_{\mu}$ the value of the color coding 
is different - see the upper and lower legend. The contours where $N_{tot}=1$ 
and $N_{\mu}=1$ are also shown by solid and dashed lines, respectively. The other 
energy fractions are taken as constants, $\ep_{e}=0.1$ and $\ep_{B}=0.01$. The 
IceCube null result favors the yellow and blue regions, e.g. a high $\eta$. } 
\end{figure}

\begin{figure} \includegraphics[scale=0.8]{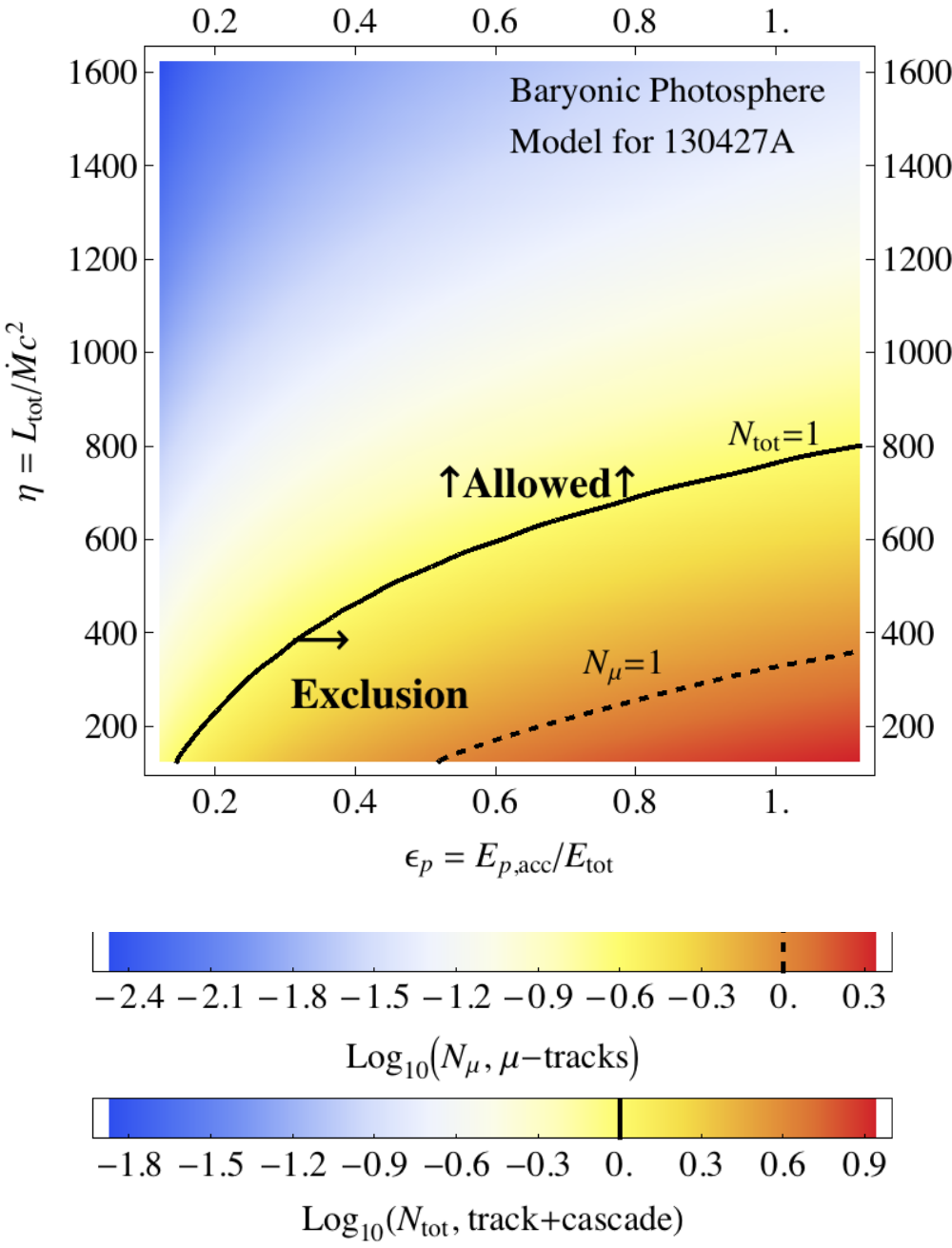} \caption{\label{fig:F3} 
The expected neutrino event number in IceCube based on the baryonic photospheric
model for GRB 130427A. The conventions are the same as those in Fig.[\ref{fig:F2}]. 
A high $\eta$ (or $\G$) is favored by the null result in IceCube.} 
\end{figure}

\begin{figure} \includegraphics[scale=0.8]{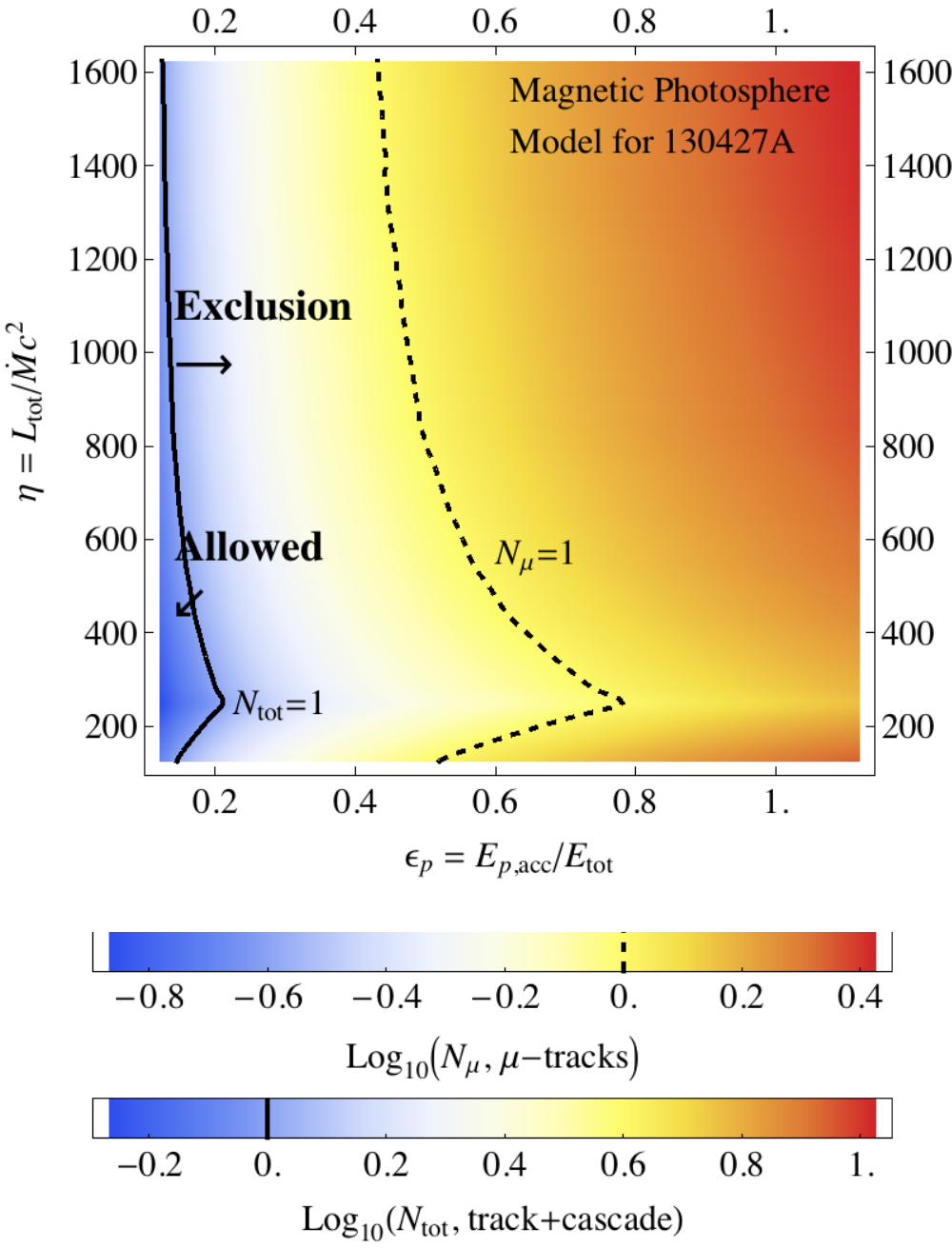} \caption{\label{fig:F4} 
The expected neutrino event number in IceCube based on the magnetic photospheric
model for GRB 130427A. The conventions are the same as those in
Fig.[\ref{fig:F2}]. The result is insensitive to $\eta$ (see \S III for an
explanation.). The null result in IceCube favors the region where
$\ep_{p}<0.1\sim0.2$, roughly independent of $\eta$ or $\G$.} 
\end{figure}

The model-independent expected number of neutrino events $N_{tot}$, including all types 
(track and cascade events) and flavors, are shown for this GRB in Fig.[\ref{fig:F1}]. 
The neutrino flux $\phi$ in the observer frame 
is computed using as input the two free parameters $R_{d}$ 
and $\G$ on a $100\times100$ grid, and this is then integrated 
with the IceCube effective area profile at the source incident angle $dec=-27$ degrees 
to obtain $N_{tot}$, from an energy range $10^{2}\le E \le10^{9}$ GeV which is 
sufficiently broad to cover the entire neutrino spectra discussed in this paper.
The value of $N_{tot}$ is represented by color. Since
$N_{tot}\propto\ep_{p}$ by eqn. \ref{eq:phiA} and \ref{eq:phiB}, we show five
contours where $N_{tot}=1$ for different $\ep_{p}/\ep_{e}$ values. The allowed region
in the $R_{d}-\G$ parameter space favors a moderately high value of $R_{d}$ and
$\G$ for small $\ep_{p}$; for high proton to lepton energy ratio , e.g.
$\ep_{p}/\ep_{e}=10$, only large $\G$ and $R_{d}$ values are admitted (blue, or
top right region of the figure). This is consistent with the nature of the $p\g$
interactions and the photon spectrum of this GRB, for two reasons: 
a) The wind comoving frame photon energy density
$u_{\g}=L/4\pi{R_{d}}^{2}\G^{2}c$ decreases rapidly with $R_{d}$ and $\G$; a
low $u_{\g}$ suppresses $f_{\pi}$ and thus the final neutrino flux. b) The
neutrino break energy, where the neutrino flux contributes most to the final
$N_{tot}$ in IceCube, is proportional to $\G^{2}$. A higher $E_{\nu{b}}$ is
associated with a smaller contribution (due to the effective area function and
the photon indices for this GRB). 

\section{Constraints on the internal shock, baryonic photosphere and magnetic 
photospheric models} 
\label{sec:3} 

In this section we discuss these three scenarios, labeled IS, BPH and MPH. 
After specifying any one of them, $R_{d}$ is derivable from $\G$ and the
other parameters, and hence one degree of freedom is eliminated from the
parameter space.  For the IS model, semi-relativistic shocks are formed by the
collision of two shells of different velocities. The dissipation radius is estimated 
as $R_{d,is}\sim \G^{2}c\Delta{t}=0.27\times10^{13}\G_{300}^{2}\Delta{t_{ms}}$ cm, 
where $\Delta{t}=10^{-3}\Delta{t_{ms}}$ s is the variability time scale in
milliseconds and $\G$ is the averaged bulk Lorentz factor of the two shells. For 
the two photospheric models, the dissipation is assumed to take place at the 
photosphere where the optical depth for a photon to scatter off an electron is unity 
$\tau_{\g{e}}=1$. Depending on whether the jet is dominated by the kinetic energy of 
the baryons or by the toroidal magnetic energy, the photospheric scenario is either
a baryonic photospheric (BPH) or  a magnetic photospheric (MPH) type. For the magnetic 
type, the fast rotating central engine (a black hole or magnetar) can lead to a highly 
magnetized outflow which is initially Poynting dominated with a sub-dominated baryonic 
load. If the magnetic field is striped , carrying alternating polarity and the jet is 
roughly one dimensional, the bulk acceleration of the jet 
is approximately $\G(r)=(r/r_{0})^{1/3}$ until a saturation radius
$r_{sat}=r_{0}\eta^{3}$, where $r_{0}=10^{7}R_{7}$ cm is the base radius of the
jet and $\eta=L_{tot}/\dot{M}c^{2}$ is the baryon load portion. Around the photosphere 
region
\footnote{While in the ICMART model by \citep{Zhang+11icmart}, the magnetic dissipation
is determined by the variability timescale, which resembles the internal shock model.}
, we assume a major fraction of the jet energy (whether baryonic or magnetic)
 is dissipated, leading to proton
acceleration, resulting in a proton spectrum 
$dN_{p}/dE\propto{E}^{-2}$ similar to that expected from a Fermi process, as is the 
case with the internal shock model, e.g. see \citep{Murase+08photonu,Wang+09photonu,
Drury+12}.
For most reasonable parameters, in the magnetic 
MPH model we have $\G \lesssim \eta$ where the jet is still in the acceleration phase 
at the dissipation radius. On the other hand, for the BPH and IS models, the dissipation 
radius almost always occurs outside the saturation radius, namely $\G=\eta$. The
determination of $R_d$ is more complicated for photosphere models than it is for the 
IS models, due to many factors (e.g. \citep{Vurm+11phot,Uzdensky+11magrecon}.)
For example, the jet may be contaminated by the electron
positron pairs which will substantially increase $\tau_{\g{e}}$ and $R_{phot}$.
A more realistic consideration should also include the fact that the
dissipation can start from the sub-photosphere all the way out, until the jet
is saturated. The magnetic field configuration is also complex (e.g. the
geometry of the layers, the reconnection rates etc) which will eventually
affect the proton and neutrino spectrum. 
Although a multi-zone simulation is
beyond the scope of this paper, it would be desirable in order to increase the 
precision of the neutrino spectrum.

In this section, we have used a calculation scheme similar to that in
\citep{Gao+12photnu} 
\footnote{The code is updated to a parallel version to 
allow the fast computation of a large region parameter space, for this paper.}.  
We make the same assumptions suitable for a one-zone calculation, but the 
consideration on the micro-processes is here more complete, compared to that 
of \S II, and is is based on a numerical code.
The pionization efficiency is obtained by calculating
all the leading order processes, e.g. $p\g$ (Delta-resonance and multi-pion
productions), Bethe-Heitler, pp, synchrotron, inverse Compton and adiabatic
losses. The cooling of the secondary charged particles via synchrotron and
inverse Compton, and the energy distribution functions for the neutrinos from
pion and muon decay are also included. The expected neutrino events detectable 
in IceCube are plotted in
Fig.[\ref{fig:F2},\ref{fig:F3},\ref{fig:F4}], for the three models separately.
The asymptotic Lorentz factor $\eta$ (instead of $\G$ at the dissipation radius) 
and the ratio of accelerated protons to electrons $\ep_{p}/\ep_{e}$ are the two 
input free parameters. The calculation is performed on a $50\times60$ grid for this
parameter space for each model. The resultant neutrino event number, $N_{\mu}$
(tracks only) and $N_{tot}$ (track+cascade) , are represented by colors. We
also show the contours in each figure where $N_{\mu}=1$ and $N_{tot}=1$.  

\noindent
{\it Constraints on IS models  (Fig.[\ref{fig:F2}])}: Here we have used a variability
timescale $\Delta{t}=1$ ms. This corresponds to a minimal dissipation radius
$R_{d}\sim2.7\times10^{13}\G_{300}^{2}$ cm which is optimally advantageous for neutrino
production. A higher value of $\Delta t$ would increase $R_{d}$ and decrease $f_{\pi}$ 
and $\phi$ from those shown in Fig.[\ref{fig:F1}].  Thus, the constraints should be 
considered looser when using a higher value of $\Delta{t}$. The results suggest that
$\ep_{p}/\ep_{e}$ values from 1 to 10 can all be admitted, but for $\ep_{p}/\ep_{e}>5$, 
$\G>600$ is required in order not to violate the IceCube null result. We note also
that $N_{tot}$ and $N_{\mu}$ increase when $\eta=\G$ is lowered, but their values
saturate at about $\eta=300$ and then decrease when $\eta<300$.  
For $\eta\lesssim300$, $f_{\pi}$ reaches unity and protons lose almost all their 
energy to pions. A smaller radius is associated with a higher energy density $u_{B}$
and $u_{\g}$, which causes the charged pions and muons to cool faster before they
decay to produce neutrinos. A smaller $\G$ value gives a smaller $E_{\nu{b}}$
value, for which IceCube has a smaller effective area. 

\vspace{2 mm}

\noindent
{\it Constraints on the BPH model (Fig.[\ref{fig:F3}])}: The photospheric radius is
estimated as $R_{phot}\approx\sigma_{T}L_{tot}/4\pi\G^{3}m_{p}c^{3}$. The
result coincidentally resembles the IS model with $\Delta t=1$ ms. However, at low $\G$ 
values, $R_{phot}$ rapidly increases, which is different from the $R_{d}\propto\G^{2}$ 
behavior of the IS model. On the contrary, only at high $\eta$ values does the magnetic
field begin to cool the charged secondaries significantly, leading to a suppressed neutrino 
spectrum.

\vspace{2 mm}

\noindent
{\it Constraints on the MPH model (Fig.[\ref{fig:F4}])}: The most interesting
constraints are obtained for the MPH model. Due to the nature of the magnetic
jet, the photosphere, if one neglects the effects of $e^{\pm}$ formation 
\citep{Veres+12mag}, should occur in the acceleration phase for the likely
parameter values. Even if the jet is initially loaded with a small amount 
of baryons (a high $\eta$ value), $\G$ at $R_{phot}$ is roughly a constant value
$150\le\G\le200$ for most $\eta$ choices. This fact is also revealed in
Fig.[\ref{fig:F4}] by the contours being almost parallel to the $\G$-axis.
The magnetic photospheric radius is larger than the $R_{d}$ computed for the IS model
with $\Delta{t}=1$ ms, but it is not large enough to suppress $f_{\pi}$ much below  
unity. The secondaries suffer somewhat less cooling from synchrotron and IC than in 
the IS case considered. A somewhat lower bulk Lorentz factor \footnote{However, $E_{\nu{b}}$ 
is also lower, which decreases the event number in IceCube.} is advantageous for
neutrino production. Therefore, the MPH model generally has an equal or higher
neutrino efficiency than the IS and BPH model. Although the result is insensitive to 
$\eta$, a relatively stringent constraint on $\ep_{p}/\ep_{e} \le 1\sim2$ is obtained 
for this burst, independent of $\eta$ or $\G$. For very low values of $\eta$, there is 
also a ``saturation effect" for reasons similar as in the IS case. 

\section{Discussion} 
\label{sec:4}

We have discussed the implications of the non-detection by IceCube in the gamma-ray 
burst GRB130427A. Using the results of the electromagnetic observations of this 
burst, we have analyzed the implications of this neutrino null-result 
for constraining the physical parameters of this burst. Using first a simplified
analysis which is independent of specific GRB models,  we find that the null-result
implies a simple inverse relationship between the bulk Lorentz factor $\G$ at 
the dissipation radius $R_d$ and this radius, as a function of the relativistic
proton to electron ratio $\ep_{p}/\ep_{e}$ (Fig.[\ref{fig:F1}]), which suggests
values of $\G \gtrsim 500$ and $R_d \gtrsim 10^{14}$ cm. We then performed 
more detailed numerical calculations for three different specific GRB models,
the internal shock (IS), baryonic photosphere (BPH) and magnetized photosphere
(MPH) models. We find that the IS model (Fig.[\ref{fig:F2}]) with the shortest 
variability time and the highest neutrino luminosity is able to comply with 
the null-result constraint if its bulk Lorentz factor $\G=\eta \gtrsim 400-600$, 
depending on $\ep_{p}/\ep_{e}$, a fairly modest constraint. Longer variability times 
only weaken the constraint.  For the baryonic photosphere BPH model (Fig.[\ref{fig:F3}]) 
the constraint for compliance is comparable, $\G=\eta \gtrsim 600-700$ depending on
$\ep_{p}/\ep_{e}$. The most restrictive constraint is for the magnetic photosphere MPH
model of Fig.[\ref{fig:F4}]. Here it is found that, for this burst GRB130427A, the
null result implies an allowed value of $\ep_{p}/\ep_{e} \lesssim 1-2$, almost 
independently of the asymptotic bulk Lorentz factor $\eta$. More careful 
calculations of all three types of models will clearly be required for reaching
firm conclusions, but based on the above considerations, the generic internal shock
and baryonic photosphere models are not significantly constrained by the lack of
observed neutrinos.

\vspace{4 mm}
 
We are grateful to the useful comments by the anonymous referee, and 
 NASA NNX13AH50G and JSPS for partial support.

\bibliographystyle{apj}

\end{document}